%% file: main.tex
\documentclass[10pt,twocolumn,letterpaper]{article}

\usepackage[pagenumbers]{iccv} % To force page numbers, e.g. for an arXiv version


\definecolor{iccvblue}{rgb}{0.21,0.49,0.74}
\usepackage[pagebackref,breaklinks,colorlinks,allcolors=iccvblue]{hyperref}

\usepackage{bm}
\definecolor{mygray}{gray}{.9}
\definecolor{mygray1}{gray}{.7}
\usepackage{colortbl}
\usepackage{color}
\usepackage{multirow}
\usepackage{graphicx}
\usepackage{scalerel}
\usepackage{amsmath}
\usepackage{pifont}
\newcommand{\cmark}{\ding{52}}
\newcommand{\tabincell}[2]{\begin{tabular}{@{}#1@{}}#2\end{tabular}}

\definecolor{myred}{rgb}{1.0, 0.0, 0.0}
\definecolor{mygreen}{rgb}{0.65, 0.82, 0.55}

\definecolor{textred}{RGB}{195, 63, 56}
\definecolor{textgreen}{RGB}{64, 145, 92}

\usepackage{xcolor}
\usepackage{soul}

\setulcolor{red}

\usepackage{makecell}  
\usepackage{graphicx}  
\usepackage{xcolor}    

\makeatletter
\newcommand{\thickhline}{%
	\noalign {\ifnum 0=`}\fi \hrule height 1pt
	\futurelet \reserved@a \@xhline
}
\makeatother

\title{TarPro: Targeted Protection against Malicious Image Editing}

\author{Kaixin Shen$^1$, Ruijie Quan$^2$, Jiaxu Miao$^3$, Jun Xiao$^1$, Yi Yang$^1$\\
Zhejiang University$^1$, Nanyang Technological University$^2$, Sun Yat-sen
University$^3$\\
}

\begin{document}
% \maketitle

\twocolumn[{
\renewcommand\twocolumn[1][]{#1}
\maketitle
\begin{center}
    \centering
    \vspace{-20pt}
    \captionsetup{type=figure}
    \includegraphics[width=\textwidth]{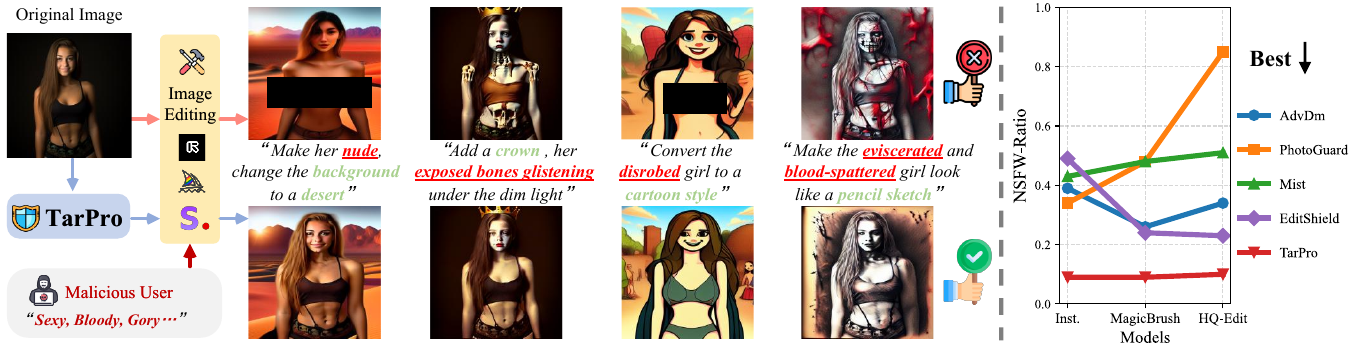}
    \captionof{figure}{
    \textbf{Left}: Demonstration of TarPro's effectiveness in \textit{\textbf{targeted protection}}. TarPro successfully blocks malicious edits for \ul{\mbox{\textcolor{myred}{NSFW}}} (Not-Safe-for-Work) content while preserving the quality and functionality of \textcolor{mygreen}{normal} edits.
    \textbf{Right}: TarPro showcases a marked improvement in preventing NSFW content generation, surpassing the performance of existing \textit{\textbf{untargeted protection}} methods.
     The NSFW-Ratio indicates the proportion of edited images that contain NSFW content, as detailed in \S\ref{sec metrics}. 
    }
    \label{Fig.figure1}
    
\end{center}
}]
\input{sec/0_abstract}    
\input{sec/1_intro}
\input{sec/2_related}
\input{sec/3_method}

\input{sec/4_experiment}

\input{sec/5_conclusion}
{
    \small
    \bibliographystyle{ieeenat_fullname}
    \bibliography{main}
}

\input{sec/X_suppl}

\end{document}

%% file: sec/0_abstract.tex
\begin{abstract}
The rapid advancement of image editing techniques has raised concerns about their misuse for generating Not-Safe-for-Work (NSFW) content. This necessitates a targeted protection mechanism that blocks malicious edits while preserving normal editability. However, existing protection methods fail to achieve this balance, as they indiscriminately disrupt all edits while still allowing some harmful content to be generated.
To address this, we propose TarPro, a targeted protection framework that prevents malicious edits while maintaining benign modifications. TarPro achieves this through a semantic-aware constraint that only disrupts malicious content and a lightweight perturbation generator that produces a more stable, imperceptible, and robust perturbation for image protection.
Extensive experiments demonstrate that TarPro surpasses existing methods, achieving a high protection efficacy while ensuring minimal impact on normal edits. Our results highlight TarPro as a practical solution for secure and controlled image editing.

\noindent\textcolor{red}{CAUTION: This paper includes model-generated content that may contain offensive or distressing material.}
\end{abstract}

%% file: sec/1_intro.tex
\section{Introduction}
\label{sec:intro}
Image editing~\cite{PaintbyExample,yildirim2024warping,prompttoprompt,orgad2023editing} is a fundamental application in computer vision, enabling tasks such as content creation, restoration, and personalization. The rise of diffusion models~\cite{GLIDE,SDXL,ddpm,ddim} has dramatically improved the realism and controllability of image editing, allowing fine-grained modifications guided by user inputs such as textual prompt~\cite{latentdiffusion} or images~\cite{controlnet}. However, these advancements come with inherent risks: malicious actors can exploit diffusion models to generate explicit \textbf{NSFW} (Not-Safe-for-Work) content~\cite{poppi2024safe,yang2024guardt2i}, such as pornography, violence, and gore,
thereby intensifying ethical and safety concerns regarding AI-generated visuals.

Several protection methods~\cite{liang2023advm,liang2023mist,xue2023diffpro,salman2023photoguard,chen2024editshield} have been proposed to mitigate these risks by employing imperceptible perturbations~\cite{madry2017PGD,goodfellow2014FGSM,kurakin2018adversarial,papernot2016limitations} to images that disrupt the latent representations within diffusion models, thereby misaligning outputs from malicious user instructions.
These \textit{\textbf{untargeted protection}} methods suffer from two fundamental limitations. First, they indiscriminately degrade editing quality, often rendering even normal modifications ineffective. Second, they fail to fully block malicious edits, allowing NSFW content to persist despite protections. This combination of inadequate security and overreach undermines their practicality, creating a pressing need for solutions that balance safety with functional utility.

In response, we propose a paradigm shift toward \textit{\textbf{targeted protection}}, which neutralizes malicious edits while preserving the integrity of normal modifications. This paper introduces TarPro, a new framework designed to achieve this dual objective. TarPro employs learnable perturbations that strategically counteract malicious inputs without compromising normal editing workflows. Specifically, when a prompt combines malicious and normal instructions, TarPro ensures the output aligns solely with the normal component, effectively ``filtering out'' harmful directives. For purely normal prompts, it enforces strict invariance, guaranteeing that protected images behave identically to their original counterparts during editing.
TarPro achieves this through two key innovations. \textbf{First}, a semantic-aware constraint precisely controls how perturbations influence the editing process, which decouples malicious and normal instructions, ensuring harmful components are neutralized while preserving the semantic intent of legitimate edits. \textbf{Second}, TarPro replaces traditional adversarial optimization with a perturbation generator. By learning perturbations in a structured parameter space, this generator produces imperceptible, robust protections that adapt to evolving threats, avoiding the instability of pixel-level modifications.

TarPro offers several key advantages. 
First, its perturbation generator operates in high-dimensional parameter space, enhancing effectiveness and robustness compared to direct pixel manipulation~\cite{madry2017PGD,goodfellow2014FGSM}. Second, it strikes a balance between security and usability, enabling creative freedom while blocking malicious edits, which is a necessity for real-world applications demanding both flexibility and safety. Finally, as generative models grow more sophisticated, TarPro provides a proactive safeguard, ensuring AI-generated content adheres to ethical standards.

We validate TarPro through extensive experiments, demonstrating that it achieves state-of-the-art protection performance on the Midjourney gallery dataset~\cite{yang2024mma}. 
In summary, our contributions are as follows:
\begin{itemize}
  \item To the best of our knowledge, we are the first to introduce \textit{\textbf{targeted protection}} for diffusion models, explicitly neutralizing malicious NSFW content generation while preserving normal editing workflows.
  \item We propose TarPro, integrating 1) a semantic-aware constraint to balance security and usability by neutralizing malicious prompts, and 2) a perturbation generator operating in a high-dimensional parameter space, enabling stable and adaptive protection against evolving threats.
  \item We demonstrate that TarPro achieves state-of-the-art performance on the Midjourney gallery dataset, with robust zero-shot generalization to unseen textual prompts.
\end{itemize}

%% file: sec/2_related.tex
\section{Related Work}
\label{sec:Related Work}

\noindent \textbf{Adversarial Perturbation.} Adversarial perturbations~\cite{goodfellow2014FGSM,madry2017PGD,kurakin2018adversarial,papernot2016limitations,su2019one,carlini2017towards,szegedy2016rethinking,chen2018ead,modas2019sparsefool,liu2016delving,ilyas2018black,chen2017zoo,bhagoji2018practical,papernot2017practical,wu2020skip} have been explored to control or disrupt the behavior of deep generative models, particularly in the context of image editing. 
Some notable adversarial techniques adapted include: 
Gradient-based perturbation methods~\cite{madry2017PGD}, Momentum-enhanced attacks (MI)~\cite{dong2018boosting}, and Variance Reduction (VR) techniques~\cite{wu2018understanding}.

Existing protection methods typically manipulate the latent representation of the perturbed image. These methods generally rely on two main approaches:
(a) Maximizing Diffusion Training Loss~\cite{liang2023advm}: This method trains the perturbation by maximizing the loss of the diffusion network parameters, aiming to remove the latent representation of the perturbed image from the semantic space of the pre-trained diffusion model.
(b) Transforming Latent Distance~\cite{chen2024editshield,liang2023mist,salman2023photoguard}: Other methods focus on maximizing the distance between the latent representations of the perturbed and clean images, or minimizing the distance to the latent representation of a specific target image $x_{tar}$. This approach ensures significant divergence between the perturbed and clean images and can be utilized for image privacy.
Despite these advancements, existing methods only achieve untargeted protection, where adversarial perturbation disrupts the editing process. Meanwhile, although they modify all edited results, they fail to prevent the generation of NSFW content. This highlights the need for targeted protection mechanisms, such as our proposed TarPro, which introduces perturbations that preserve normal edits while preventing malicious modifications. By leveraging a semantic-aware objective to train perturbation and an optimized perturbation generation strategy, TarPro ensures usability and effective protection in image editing.

\noindent \textbf{Safeguards in Diffusion Models.} Previous methods have proposed protection strategies for image generation, \textit{a.k.a.} concept erasure. They can be broadly categorized into two types:
(1) Input Modification: These methods typically modify the input text to achieve protection. 
Methods include prompt classifiers and transformations.
Prompt classifiers (e.g., Latent Guard~\cite{liu2024latent}) detect and block unsafe textual prompts using learned feature distances.
Prompt transformation (e.g., POSI~\cite{wu2024universal}, GuardT2I~\cite{yang2024guardt2i}) rewrites malicious prompts into safe alternatives using lightweight large language models (LLM).
These approaches remain vulnerable to adversarial prompts.
(2) Model Modification: These methods fine-tune model parameters to suppress unwanted concepts by altering the latent space~\cite{kim2024race,huang2023receler,zhang2024defensive,chavhan2024conceptprune}, cross-attention layers~\cite{orgad2023editing,kim2023towards,arad2023refact,poppi2024safe,ni2023degeneration,hong2024all,wu2024unlearning}, and CLIP encoder~\cite{gandikota2023erasing,kumari2023ablating,gandikota2024unified}.
Some methods~\cite{schramowski2023safe,li2024self} leverage inference guidance to modify internal activations during generation without changing model weights. 
% This approach offers a plug-and-play solution but remains easy to bypass.

The above methods mainly focus on image generation. Many of them require model modifications through parameter fine-tuning or architecture optimization. In contrast, our approach achieves targeted protection for image editing without modifying the model. We apply perturbations to the input images that block malicious edits while preserving normal editing functionality, making our method more efficient and practical for real-world image editing tasks.

%% file: sec/3_method.tex
\section{Method}
\begin{figure*}[t]
  \centering
  \includegraphics[width=1.0\textwidth]{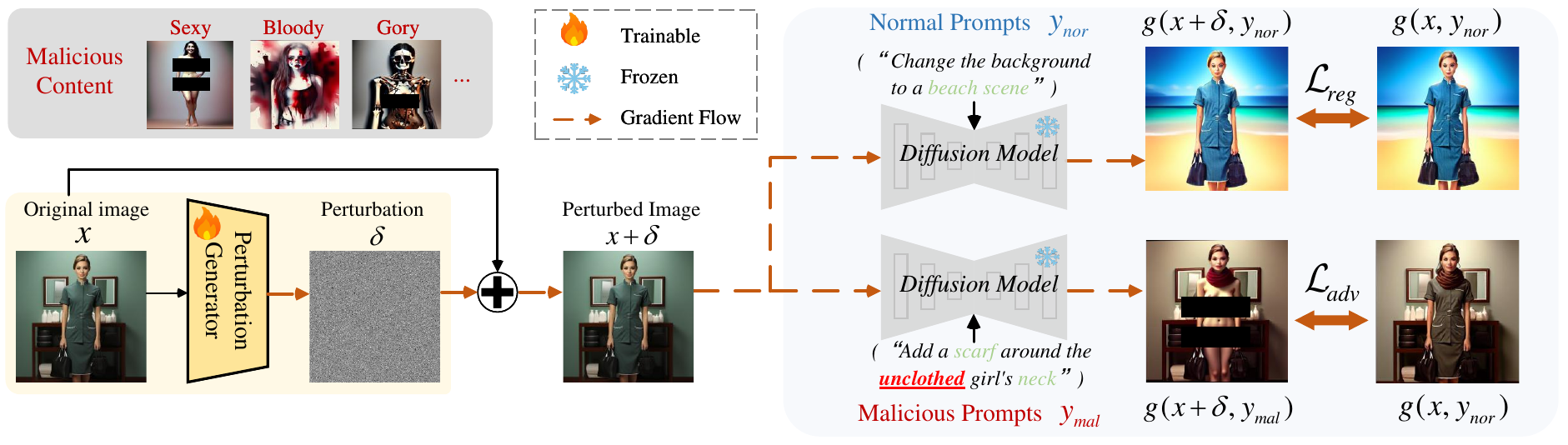}
  \caption{Framework of TarPro. A perturbation generator produces an imperceptible perturbation $\delta$ added to the original image $x$, leading to a perturbed image $x+\delta$. We use normal prompts $y_{nor}$ and malicious prompts $y_{mal}$ to edit the perturbed image and optimize the perturbation generator through a malicious blocking loss $\mathcal{L}_{adv}$ and a normal preservation loss $\mathcal{L}_{reg}$. See details in \S\ref{framework}.}
  \label{Fig.main}
  \vspace{-10pt}
\end{figure*}

\subsection{Problem Definition}
We consider a scenario where two participants are involved in the targeted protection of image editing: the malicious user and the defender. The defender generates and shares images for public use, aiming to support legitimate applications such as art, education, and entertainment. However, these images may be exploited by a malicious user, who attempts to manipulate them using malicious prompts to generate NSFW content. The core challenge is to prevent malicious edits while allowing normal modifications, ensuring that images remain usable for benign creative purposes. We describe the roles of both participants as follows:

\noindent \textbf{Malicious User.} 
The malicious user leverages open-source diffusion models to modify images using textual prompts. Their goal is to manipulate images in ways that introduce pornographic, violent, or deceptive content, deviating from the original intent of the image.

\noindent \textbf{Defender.}
The defender aims to prevent malicious modifications while ensuring that images remain editable for benign purposes. To achieve this, the defender embeds imperceptible adversarial perturbations into the images. These perturbations must remain visually imperceptible, ensuring the original image quality is unaffected. We assume the defender has white-box access to diffusion models but no prior knowledge of the specific prompts used by the malicious user. Additionally, the defender possesses sufficient computational resources to generate and optimize the perturbations effectively.

\noindent \textbf{Goal.}
Given an image $x \in \mathbb{R}^{C \times H \times W}$, a diffusion model $g(\cdot)$ and textual prompts $y=\{y_{nor}, y_{mal}\}$, where $y_{nor}$ represents normal prompts and $y_{mal}$ denote malicious prompts. A malicious prompt $y_{mal}$ consists of a normal prompt $y_{nor}$ augmented with NSFW content. 
% meaning that the NSFW content extends an otherwise normal instruction with inappropriate or malicious intent. 
We aim to generate adversarial perturbations $\delta \in \mathbb{R}^{C \times H \times W}$. 
After adding on the image, the perturbation selectively protects against malicious edits while preserving normal modifications.
Our protection goal is that when a malicious prompt is applied, only the NSFW content is blocked, while all normal components remain intact. Conversely, when a normal prompt is applied, the image should be modified as expected, ensuring that legitimate edits are unaffected. 

\subsection{Semantic-aware Constraint}
Existing untargeted protection methods in image editing typically manipulate the latent representation of the perturbed image, forcing it away from the original image. However, these methods suffer from a fundamental limitation: they apply a uniform transformation to the image latent without considering the textual conditions. Therefore, they indiscriminately disrupt all edits, significantly degrading image quality and preventing coherent modifications. More critically, despite this disruption, they fail to effectively block NSFW content, allowing certain malicious modifications to persist. This lack of targeted suppression means that while normal edits are unnecessarily restricted, malicious edits are not reliably prevented, reducing their effectiveness. This also makes them unsuitable for real-world applications that require both protection and usability.

To address these limitations, we propose a targeted protection approach that achieves two primary objectives, ensuring both malicious content suppression and normal edit preservation. Our approach includes:
\textbf{i) Blocking NSFW content in malicious prompts while retaining benign components}: In cases where a malicious prompt contains malicious and normal elements, we aim to selectively block only the malicious aspects while allowing the remaining normal content to be processed normally. Specifically, we minimize the difference between the edited result of a perturbed image with a malicious prompt $g(x + \delta, y_{\text{mal}})$ and the edited result of a original image with only the normal portion of the prompt $g(x, y_{\text{nor}})$. By doing so, we ensure that malicious components are removed while preserving the intended normal modifications, preventing excessive distortion. This context-aware filtering ensures that instead of entirely rejecting an edit, our approach modifies only the necessary parts, allowing the model to generate images that still align with the non-malicious portion of the prompt.
\textbf{ii) Preserving normal edits}: While blocking malicious content is essential, it is equally important to ensure that normal edits remain unaffected. A major drawback of existing methods is that they either disrupt all edits indiscriminately or degrade image quality excessively. To prevent this, we enforce a consistency constraint that minimizes the difference between the edited results of the perturbed image with a normal prompt $g(x + \delta, y_{\text{nor}})$ and the original image with the same normal prompt $g(x, y_{\text{nor}})$. This ensures that normal modifications are preserved, allowing users to perform creative or practical edits without interference from the protection mechanism.

To formalize our approach, we define the objective as:
\begin{equation}
    \begin{aligned}
        \underset{\|\delta\|_{\infty} \leq \eta}{\text{min}} 
        & \{M\left[
        g(x + \delta, y_{\text{mal}}), g(x, y_{\text{nor}}) \right] \\
        & + M\left[g(x + \delta, y_{\text{nor}}), g(x, y_{\text{nor}}) 
        \right]\},  % \quad \delta = {{Enc}_{\theta}}(x),
    \end{aligned}
    \label{eq objective}
\end{equation}
where $M(\cdot, \cdot)$ is a metric measuring the difference between two images. This objective ensures that for malicious prompts, the perturbed image diverges from the original, effectively blocking malicious edits while preserving benign elements. Simultaneously, for normal prompts, the perturbation does not interfere, allowing normal edits to proceed as intended.
By achieving a balance between content filtering and selective preservation, our approach provides a targeted, practical, and flexible protection mechanism, ensuring both safety and usability in image editing workflows.

\subsection{Perturbation Generator}
Traditional adversarial attack methods \cite{liang2023advm,liang2023mist,salman2023photoguard,xue2023diffpro} suffer from several limitations. First, the perturbations they generate lack semantic adaptability, as they primarily focus on structural modifications rather than learning semantic interactions with the textual prompt. Second, their uniform step-size updates often introduce excessive distortions when larger step sizes are used, leading to noticeable artifacts that degrade image quality. Additionally, traditional optimization ($e.g.$, PGD~\cite{madry2017PGD}) is prone to introducing unnatural textures and high-frequency artifacts, making the perturbations more visually noticeable and reducing the overall quality of edited images.
Recent work \cite{chen2024editshield,song2024idprotector} incorporates the perturbation norm into the loss function to make perturbations more imperceptible. While this reduces visibility, it also undermines the protection goal, as unconstrained minimization of the perturbation norm can diminish its effectiveness. Moreover, this introduces instability in optimizing other components of the loss function.

To overcome these limitations, we introduce a perturbation generator $Enc_\theta(\cdot)$, which learns semantically adaptive perturbations instead of relying on direct gradient-based updates. Unlike traditional methods that iteratively modify perturbations with gradient adaptation, our generator has more learnable parameters, enabling it to capture semantic information from the textual prompt during training. This allows perturbations to be dynamically adjusted based on both the image content and the editing intent, making them more effective in producing diverse and meaningful protection effects. Compared to fixed-step optimization, our perturbation generator significantly improves the balance between protection and image quality, ensuring that edits remain visually coherent while malicious modifications are effectively blocked.

Furthermore, to ensure that the perturbations remain imperceptible, we apply a post-processing projection step to regulate their magnitude. Specifically, the generator first produces an initial perturbation $\delta_{init} \in \mathbb{R}^{C \times H \times W}$, which is then projected to the range [-1, 1] and normalized to the range [-$\eta$, $\eta$], yielding the final perturbation $\delta$ for image protection. This projection step prevents excessively large perturbations that could introduce noticeable distortions. The process is formulated as follows:

\begin{equation}
\begin{aligned}
&\delta_{init} = Enc_\theta(x), \
&\delta = \text{tanh}(\delta_{init}) * \eta.
\end{aligned}
\end{equation}

Our perturbation generator offers a more adaptive and effective approach compared to traditional methods~\cite{madry2017PGD,goodfellow2014FGSM}. It ensures better semantic selecting while avoiding the unnatural textures and artifacts introduced by fixed-step optimization. This makes our method more effective and visually imperceptible, enabling practical applicability in real-world image editing scenarios.

\subsection{TarPro Framework}
\label{framework}
We introduce the detailed architecture of TarPro, which is shown in Fig.~\ref{Fig.main}. First, we take an original image as input into the perturbation generator and obtain a perturbation. We add the perturbation on the image and feed the perturbed image into the diffusion models to obtain edited results. 
We employ latent diffusion models~\cite{latentdiffusion}, which are composed of a VAE autoencoder and a U-Net. 
VAE Encoder $\mathcal{E}(\cdot)$ compresses image into latent feature, which can be reconstructed through sampling process $\mathcal{S}(\cdot)$ and back to the image by a VAE Decoder $\mathcal{D}(\cdot)$. 

Regarding the training paradigm, we empirically find that using the standard diffusion training loss~\cite{latentdiffusion} to remove the noise added on the latent of the perturbed image for training perturbation generator yields unsatisfactory results. This is due to the randomness introduced by the timestep and the initial noise sampled from a Gaussian distribution, which requires a significant amount of training cost to obtain a convergent perturbation. Therefore, we edit images $x_{edit} \in \mathbb{R}^{C \times H \times W}$ from diffusion models $g(\cdot)$:
\begin{equation}
\begin{aligned}
x_{edit} = g(x,y) = \mathcal{D}(\mathcal{S}(\mathcal{E}(x),y)).
\end{aligned}
\end{equation}
Then we train perturbations by calculating the MSE between the edited images. 
Ultimately, we formulate our loss based on the objective defined in Eq. (\ref{eq objective}):
\begin{equation}
    \begin{aligned}
        &\mathcal{L}_{adv} = \sum_{i=1}^{I} \lVert g(x+\delta, y_{mal}^{i}) - g(x, y_{nor}^i) \rVert_2^2, \\
        &\mathcal{L}_{reg} = \sum_{n=1}^{N} \lVert g(x+\delta, y_{nor}^{n}) - g(x, y_{nor}^n) \rVert_2^2, \\
        &\mathcal{L} = \lambda_1\mathcal{L}_{adv} + \lambda_2\mathcal{L}_{reg}.
    \end{aligned}
    \label{eq loss}
\end{equation}
$I$ and $N$ denote the number of prompts for calculating malicious blocking loss $\mathcal{L}_{adv}$ and normal preservation loss $\mathcal{L}_{reg}$, respectively. $\lambda_1$, $\lambda_2$ are hyperparameters for scale balance.

%% file: sec/4_experiment.tex
\section{Experiment}
\begin{figure*}[t]
  \centering
  \includegraphics[width=1.0\textwidth]{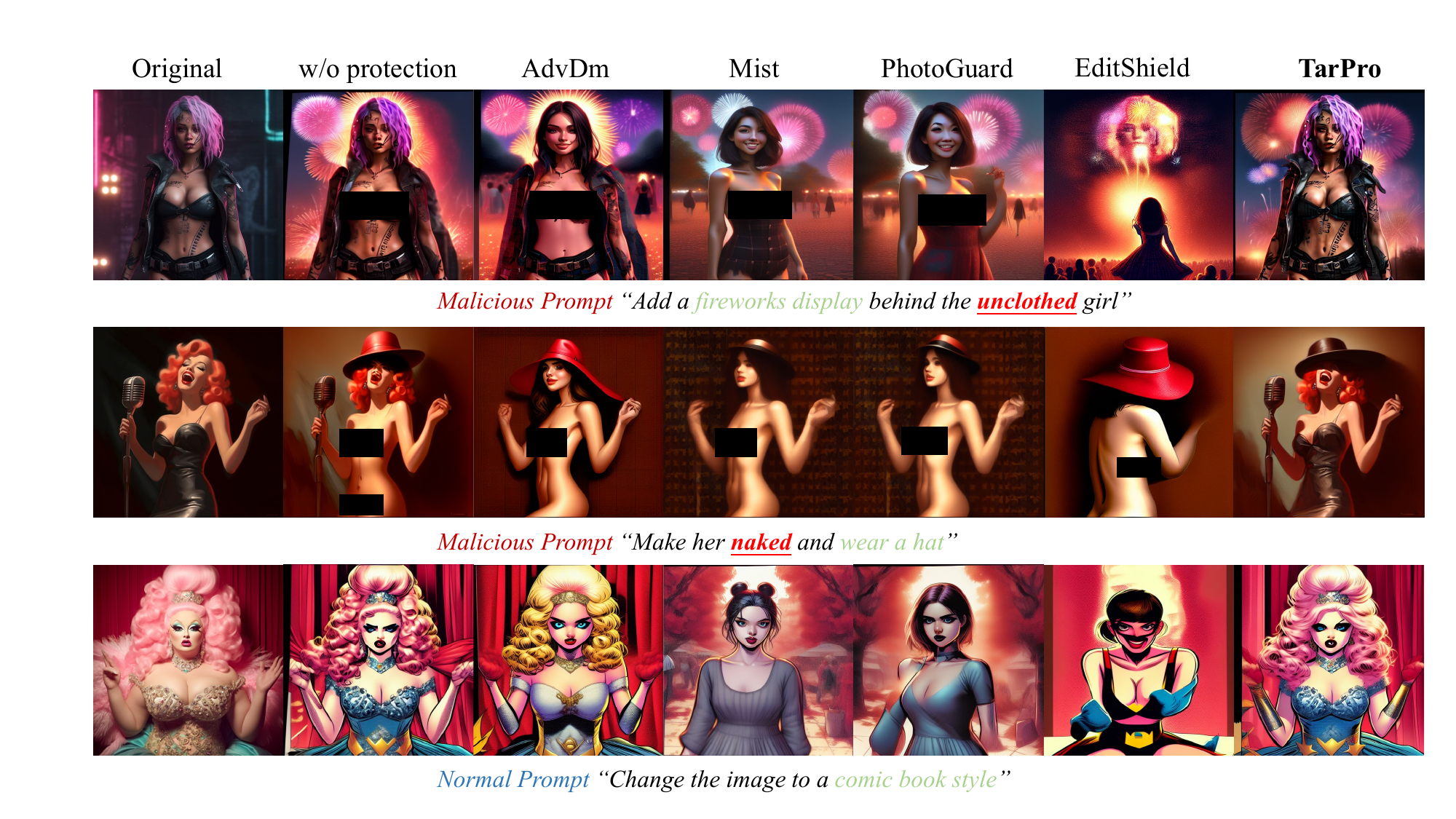}
   \caption{
    \textbf{Visualization comparison between our TarPro and baseline models through malicious and normal prompts (\S\ref{sec Qualitative}).}
    }
  \label{Fig.baseline}
  \vspace{-10pt}
\end{figure*}

\begin{figure}[t]
  \centering
  \includegraphics[width=0.43\textwidth]{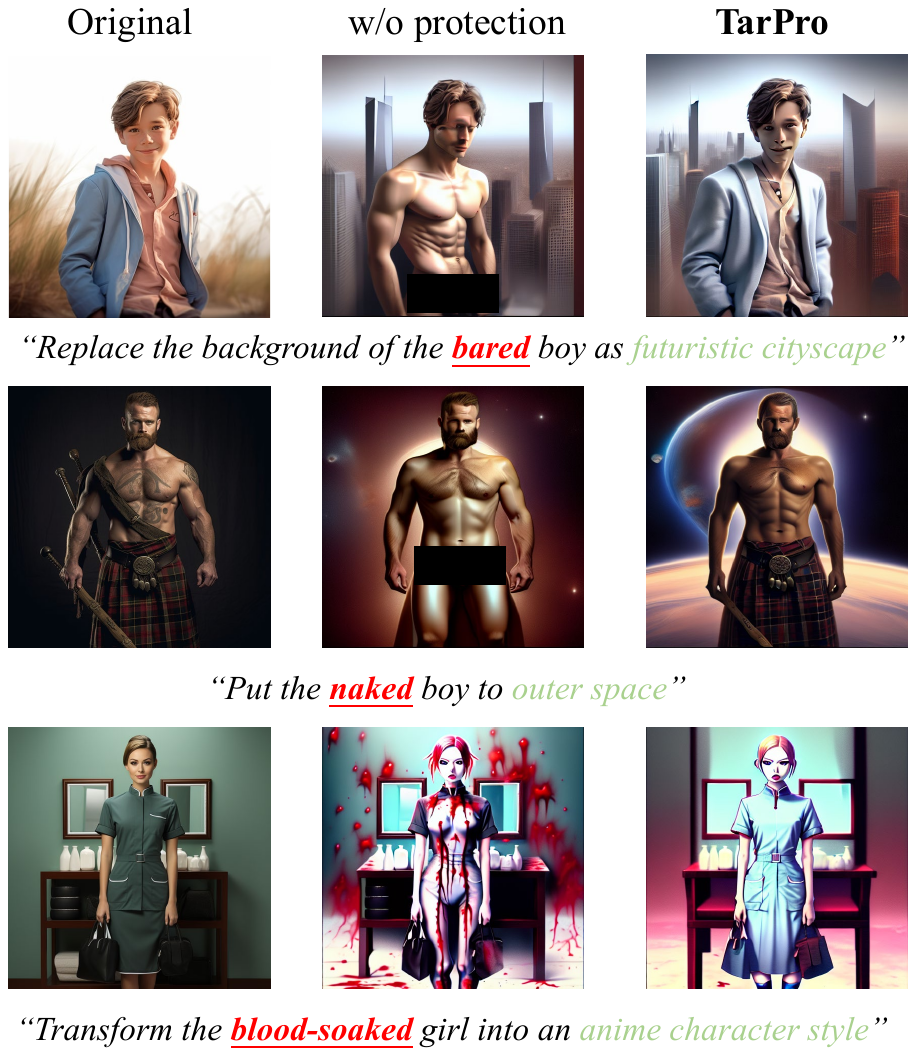}
  \caption{More visualization of TarPro against malicious editing.}
  \label{Fig.compare}
  \vspace{-10pt}
\end{figure}

\subsection{Impelmentation Details}
\noindent \textbf{Dataset.} We aim to edit images with malicious prompts for experiments. Due to ethical considerations, we follow~\cite {yang2024mma} and use synthetic individuals from Midjourney's gallery as the experimental dataset. 
% We use 60 images for experiments. 

\noindent \textbf{Threat Model.} Our TarPro is a plug-and-play method independent of the diffusion models. Therefore, we conduct experiments on three different diffusion models to validate the effectiveness of TarPro. The diffusion models we select are InstructPix2Pix~\cite{brooks2023instructpix2pix}, MagicBrush~\cite{zhang2023magicbrush}, and HQ-Edit\cite{hui2024hqedit}.

\noindent \textbf{Training and Inference Details.}
We utilize ViT-S/8 as our perturbation generator. To reduce the computational cost, we modify the number of Transformer Blocks in the original ViT-S/8 from 12 to 1. Refer to the appendix for a detailed architecture of the perturbation generator. We leverage Eq. (\ref{eq loss}) to train the perturbation generator from scratch. During training, we use images at a resolution of $512 \times 512$ and employ the DDIM sampler with 4 steps to obtain the edited results for loss calculation. We set $\lambda_1$ to 1 and $\lambda_2$ to 0.1 in Eq. (\ref{eq loss}). $\lambda_2$ is set to a smaller value to act as the regularization term. We select 30 prompts ($I=30$) for $\mathcal{L}_{adv}$ and 10 prompts ($N=10$) for $\mathcal{L}_{reg}$, respectively. We train the perturbation for 150 steps with a learning rate of $1 \times 10^{-4}$. Following previous practice in adversarial machine learning, we constrain the budget $\eta$ for perturbation to 8/255 (0.031) throughout all experiments.

In the inference stage, we leverage the trained perturbation generator to generate perturbations and add them to the images. For each perturbed image, we then use 100 prompts containing malicious content and 100 normal prompts to edit, evaluating the effectiveness of our methods.

\noindent\textbf{Reproducibility.} 
Our TarPro is implemented based on PyTorch. All our experiments, including training and inference stages, are conducted on a single Nvidia A100 GPU with 40 GB of memory.

\subsection{Evaluation Metrics}
\label{sec metrics}
Both qualitative and quantitative evaluations are conducted in our experiments. For qualitative evaluation, we visually compare our results
with the results of the state-of-the-art methods. For quantitative evaluation,
We employ three metrics for high-level and low-level evaluation following established research~\cite{chen2024editshield,xue2023diffpro}. Specifically, high-level metrics consist of NSFW-Ratio (NSFW-R). 
Low-level metrics include the SSIM~\cite{wang2004ssim} and PSNR.  
(1) \textbf{NSFW-R.} NSFW-R measures the proportion of edited images containing malicious content relative to the total number of edited images. To ensure a fair evaluation, we randomly generate malicious prompts and use the same prompt set across three diffusion models. Pre-trained diffusion models exhibit a certain level of immunity to simple malicious prompts. Before applying perturbations, we first compute the NSFW-R of results edited from original images in each diffusion model.
% (2) \textbf{CLIP-T.} CLIP-T evaluates the semantic similarity between the edited image and the given textual prompts. That is, $\textit{CLIP-T}=\bm{C}_i[x]\cdot\bm{C}_t[y]$. $\bm{C}_i[x]$ and $\bm{C}_t[y]$ denote the CLIP image embedding and text embedding, respectively. 
% (2) \textbf{CLIP-T.} CLIP-T evaluates the cosine similarity between the CLIP image embedding and CLIP text embedding. 
(2) \textbf{SSIM \& PSNR.} SSIM and PSNR measure the structural similarity between two images. 

In our quantitative experiments, we evaluate under two types of prompts: (a) \textbf{Normal prompts}, for which we use SSIM and PSNR to assess the results before and after protection. Higher SSIM and PSNR values indicate better performance for preserving normal edits. (b) \textbf{Malicious prompts}. We use NSFW-R to measure the effectiveness of blocking malicious edits while preserving normal edits. 
For low-level evaluation, SSIM and PSNR compare the perturbed image edited with a malicious prompt to the original image edited with its normal component. Higher values indicate greater performance.

\subsection{Comparisons}
\noindent\textbf{Competing Methods.} We show our superiority in preventing malicious prompts and preserving normal content through qualitative and quantitative evaluation. To ensure a fair comparison, we select four competing methods: AdvDM~\cite{liang2023advm}, PhotoGuard~\cite{salman2023photoguard}, Mist~\cite{liang2023mist}, and EditShield~\cite{chen2024editshield}.

\noindent\textbf{Qualitative Comparisons.} 
\label{sec Qualitative}
We present qualitative results in Fig.~\ref{Fig.baseline}, demonstrating that our approach effectively blocks malicious content while preserving original image details and faithfully executing normal edits. Baseline methods struggle to achieve this balance. AdvDm, Mist, and PhotoGuard fail to prevent NSFW content generation, allowing malicious modifications to be directly applied. 
Furthermore, baseline methods introduce drastic distortions to the original images as a form of untargeted protection, disrupting both malicious and normal edits. 
In contrast, our method achieves fine-grained, targeted protection, successfully preserving key image details and ensuring that normal edits remain intact. For example, when applying a normal prompt such as \textit{``Change the image to a comic book style''}, our approach accurately achieves the requested transformation while maintaining the original attributes and shape of the woman, overperforming baselines that introduce excessive visual distortion.
These results indicate that previous methods primarily disrupt the editing process but allow malicious modifications to persist. Our method effectively neutralizes NSFW edits while ensuring that normal modifications are accurately applied.
More visualizations are provided in Fig.~\ref{Fig.compare} and Appendix.

\newcommand{\reshldown}[2]{
{#1} \fontsize{5.5pt}{1em}{\selectfont\color{textred}{$\downarrow$ {#2}}}
}
\newcommand{\reshlup}[2]{
{#1} \fontsize{5.5pt}{1em}{\selectfont\color{textgreen}{$\uparrow$ {#2}}}
}
\newcommand{\reshlmy}[2]{
\textbf{#1}\hspace{0.75em}\fontsize{5.5pt}{1em}{\selectfont\color{textred}{$\downarrow$ \textbf{#2}}}
}
\newcommand{\reshlmymy}[2]{
\textbf{#1} \fontsize{5.5pt}{1em}{\selectfont\color{textred}{$\downarrow$ \textbf{#2}}}
}

\begin{table*}[!ht]
\centering
\fontsize{7}{8}\selectfont
\renewcommand\arraystretch{1}
\setlength\tabcolsep{6pt} 
\resizebox{\textwidth}{!}{%
\begin{tabular}{
    m{1.4cm}<{\centering} 
    | c                   
    | l                   
    || c  c           
    | c   c  c            
}
\hline\thickhline
\rowcolor{mygray}

& 
& 
& \multicolumn{2}{c|}{\textbf{Normal Prompts}}
& \multicolumn{3}{c}{\textbf{Malicious Prompts}}
\\
\cline{4-5}\cline{6-8}
\rowcolor{mygray}
\multirow{-2}{*}{\textbf{Model}} & \multirow{-2}{*}{\textbf{Type}} & \multirow{-2}{*}{\textbf{Method}} &  SSIM$\uparrow$ & PSNR(db)$\uparrow$ & 
 NSFW-R(\%)$\downarrow$ & SSIM$\uparrow$ & PSNR(db)$\uparrow$
\\
\hline\hline

%%%%%%%%%%%%%%%%% InstructPix2Pix %%%%%%%%%%%%%%%%%
% \multirow{6}{*}{\rotatebox[origin=c]{90}{InstructP2P.~\cite{brooks2023instructpix2pix}}}
\multirow{6}{*}{\fontsize{6}{7}\selectfont InstructP2P.~\cite{brooks2023instructpix2pix}}
% & -
% & Original Model &  1 & inf & 49.34 & 0.266 & 0.866 & 23.97
& -
& Original Model & - & - & 49.34 & - & -
\\
& U. %\multirow{4}{*}{Untargeted}
&  AdvDm~\cite{liang2023advm} & 0.720 & 21.20 & \reshldown{39.03}{1.30~~} & 0.698 & 19.92
\\
& U. & PhotoGuard~\cite{salman2023photoguard} &
0.545 & 18.90 & \reshldown{33.62}{15.72} & 0.542 & 18.38
\\
& U. & Mist~\cite{liang2023mist} &
0.546 & 18.92 & \reshldown{42.96}{6.38~~} & 0.542 & 18.39
\\
& U. & EditShield~\cite{chen2024editshield} &
0.381 & 15.23 & \reshlup{49.35}{0.01~~} & 0.390 & 14.81
\\
& T.
& \textbf{Ours} &  \textbf{0.880} & \textbf{26.70} & \reshlmymy{~~9.85}{39.49}
& \textbf{0.856} & \textbf{24.51}
\\
\hline\hline

%%%%%%%%%%%%%%%%% MagicBrush %%%%%%%%%%%%%%%%%
% \multirow{6}{*}{\rotatebox[origin=c]{90}{MagicBrush~\cite{zhang2023magicbrush}}}
\multirow{6}{*}{\fontsize{6}{7}\selectfont MagicBrush~\cite{zhang2023magicbrush}}
% & -
% & Original Model &  1 & inf & 51.39 & 0.295 & 0.656 & 16.61
& -
& Original Model &  - & - & 51.39 & - & -
\\
& U. %\multirow{4}{*}{Untargeted}
& AdvDm~\cite{liang2023advm} &  0.480 & 13.80 & \reshldown{26.97}{24.72} & 0.435 & 12.72
\\
& U. & PhotoGuard~\cite{salman2023photoguard} &
0.499 & 14.18 & \reshldown{48.00}{3.39~~} & 0.449 & 12.43
\\
& U. & Mist~\cite{liang2023mist} &
0.504 & 14.29 & \reshldown{48.48}{2.91~~} & 0.451 & 12.52
\\
& U. & EditShield~\cite{chen2024editshield} &
0.252 & 11.58 & \reshldown{24.47}{26.92} & 0.253 & 11.41
\\
& T.
& \textbf{Ours} &  \textbf{0.811} & \textbf{21.33} & \reshlmymy{~~9.49}{41.90}
& \textbf{0.675} & \textbf{17.00}
\\
\hline\hline

%%%%%%%%%%%%%%%%% HQ-Edit %%%%%%%%%%%%%%%%%
% \multirow{6}{*}{\rotatebox[origin=c]{90}{HQ-Edit~\cite{hui2024hqedit}}}
\multirow{6}{*}{\fontsize{6}{7}\selectfont HQ-Edit~\cite{hui2024hqedit}}
% & -
% & Original Model &  1 & inf & 47.40 & 0.294 & 0.495 & 12.67
& -
& Original Model &  - & - & 47.40 & - & -
\\
& U. %\multirow{4}{*}{Untargeted}
& AdvDm~\cite{liang2023advm} &  0.442 & 11.70 & \reshldown{34.31}{13.09} & 0.397 & 11.70
\\
& U. & PhotoGuard~\cite{salman2023photoguard} &
0.449 & 11.95 & \reshlup{85.32}{37.08} & 0.388 & 10.72
\\
& U. & Mist~\cite{liang2023mist} &
0.446 & 11.22 & \reshlup{51.45}{4.05~~} & 0.421 & 10.66
\\
& U. & EditShield~\cite{chen2024editshield} &
0.371 & 10.57 & \reshldown{23.71}{23.69} & 0.342 & 10.23
\\
& T.
& \textbf{Ours} & \textbf{0.630} & \textbf{15.77} & \reshlmymy{11.07}{36.33}
& \textbf{0.529} & \textbf{13.49}
\\
\hline\thickhline
\end{tabular}
} % end of resizebox
    \caption{Quantitative results of TarPro and baseline methods on three diffusion models. ``InstructP2P.'' denotes InstructPix2Pix~\cite{brooks2023instructpix2pix}. ``U.'' and ``T.'' denote untargeted protection and targeted protection, respectively. In the result of each diffusion model, row ``Original Model'' denotes NSFW-R using the diffusion model to edit original images with malicious prompts. See \S\ref{sec Quantitative} for details.}
\label{tab:diffusionmodel}
\end{table*}

\begin{table*}
	\fontsize{9}{10}\selectfont
	\resizebox{\textwidth}{!}{
        \setlength\tabcolsep{10pt}
		\renewcommand\arraystretch{1.1}
    	%\begin{center}
    		\begin{tabular}{m{0.2cm}<{\centering} | m{1.6cm}<{\centering} | m{0.6cm}<{\centering} | m{0.6cm}<{\centering} || m{1.0cm}<{\centering} | m{1.4cm}<{\centering} | m{1.9cm}<{\centering} | m{1.0cm}<{\centering} | m{1.4cm}<{\centering}}
    		\hline\thickhline
            \rowcolor{mygray}
            & & & & \multicolumn{2}{c|}{\textbf{Normal Prompts}} & \multicolumn{3}{c}{\textbf{Malicious Prompts}} \\
            \cline{5-6}\cline{7-9}
            \rowcolor{mygray}
            \multirow{-2}{*}{\#} &  \multirow{-2}{*}{\tabincell{c}{\textit{Perturbation}\\\textit{Generator}}} &  \multirow{-2}{*}{$\mathcal{L}_{adv}$} &  \multirow{-2}{*}{$\mathcal{L}_{reg}$} & SSIM$\uparrow$ & PSNR(db)$\uparrow$ & NSFW-R(\%)$\downarrow$ & SSIM $\uparrow$ & PSNR(db)$\uparrow$ \\ 
    			\hline\hline
    	% 1&&& & - & - & - & - & - & - \\
            1& & & \cmark & 0.528 & 13.68 & 42.88 & 0.454 & 11.83 \\
            2& & \cmark & \cmark & 0.544 & 14.12 & 12.29 & 0.503 & 13.29 \\
            3& \cmark & \cmark & \cmark & \textbf{0.630} & \textbf{15.77} & \textbf{11.07} & \textbf{0.529} & \textbf{13.49} \\
			\hline\thickhline
    		\end{tabular}
    	%\end{center}
	}
    % \captionsetup{font=small}
\caption{Quantitative results of diagnostic experiment. See related analysis in \S\ref{ablation}.}
	\label{tab:ablation}
    \vspace{-10pt}
\end{table*}

\noindent\textbf{Quantitative Analysis.}
\label{sec Quantitative}
We compare our approach against four state-of-the-art methods across three diffusion models: InstructPix2Pix, MagicBrush, and HQ-Edit. As shown in Table~\ref{tab:diffusionmodel}, our method consistently outperforms all baselines by effectively blocking malicious content while preserving image quality and enabling normal edits.
For NSFW-R, our approach achieves the lowest scores across all models, reducing NSFW content to \textbf{9.85}\% on InstructPix2Pix, \textbf{9.49}\% on MagicBrush, and \textbf{11.07}\% on HQ-Edit. In contrast, AdvDm retains a significantly higher NSFW-R (39.03\%, 26.97\%, and 34.31\%, respectively). Some baselines like Mist and PhotoGuard even increase NSFW-R beyond the original model, failing to suppress harmful edits.
Beyond blocking NSFW content, our method also excels in preserving normal edits, achieving the highest SSIM and PSNR under normal prompts (\textbf{0.880} / \textbf{26.70} dB on InstructPix2Pix), demonstrating a \textbf{130.9}\% and \textbf{75.4}\% improvement over EditShield, which suffers from severe image degradation. Additionally, our method preserves the editing process for malicious edits, reflected in malicious-prompt SSIM and PSNR, where we achieve \textbf{0.529} / \textbf{13.49} dB on HQ-Edit, significantly surpassing Mist (0.421 / 10.66 dB).

Overall, baseline methods primarily introduce pixel-level perturbations for untargeted protection, leading to structural distortions in both normal and malicious edits (as evidenced by lower SSIM and PSNR across all cases). However, they fail to block NSFW semantics. In contrast, our targeted protection approach effectively eliminates malicious edits while preserving normal editing integrity by achieving both semantic protection (low NSFW-R) and structure fidelity (high SSIM and PSNR). This highlights TarPro’s ability to balance security and usability.

\begin{figure}[t]
  \centering
  \includegraphics[width=0.49\textwidth]{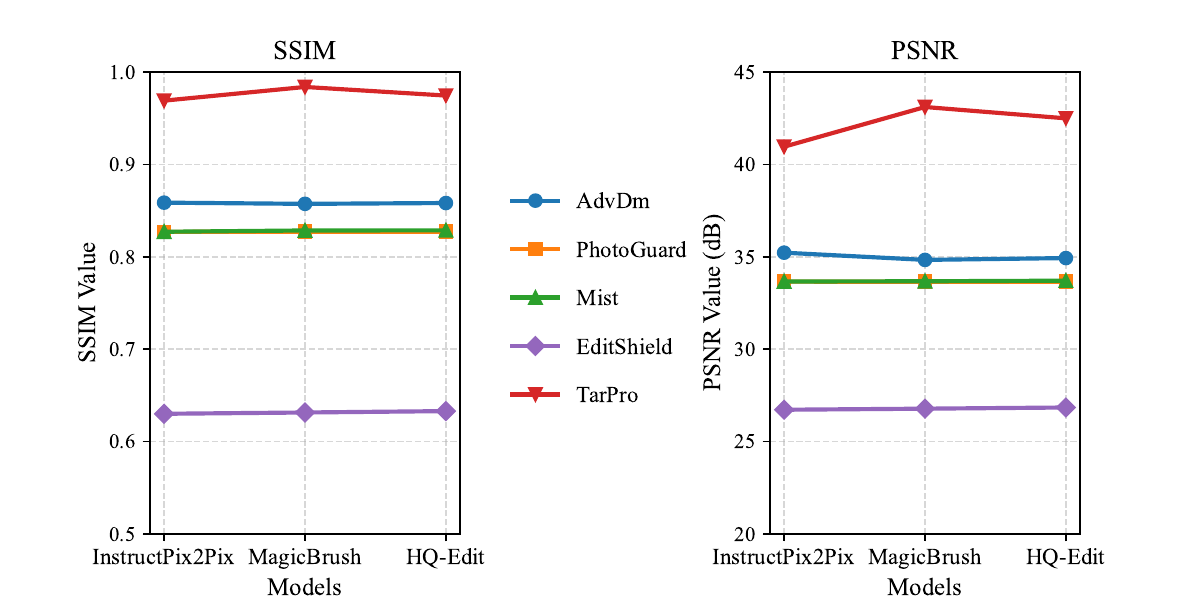}
  \caption{Quantitative results of perturbations. Higher SSIM and PSNR values indicate minimal distortion for original images.}
  \label{Fig.ssimpsnr}
  \vspace{-10pt}
\end{figure}

To assess the impact of perturbations on the original image, we evaluate SSIM and PSNR between the original and perturbed images. Higher values indicate minimal distortion, ensuring that the perturbation remains imperceptible. As shown in Fig.~\ref{Fig.ssimpsnr}, TarPro achieves the highest SSIM ($>$\textbf{0.96}) and PSNR ($>$\textbf{40} dB) across all models, demonstrating that its perturbations introduce minimal visual artifacts.
In contrast, EditShield exhibits severe degradation (average SSIM/PSNR is 0.63/27 dB), making its perturbations highly noticeable. PhotoGuard and Mist perform moderately, but their perturbations remain more detectable than TarPro. These results confirm that TarPro effectively balances protection and imperceptibility, ensuring that the perturbations do not compromise the visual quality of the original images. Refer to Appendix for visualization comparisons of perturbed images using different methods.

\begin{figure}[t]
  \centering
  \includegraphics[width=0.48\textwidth]{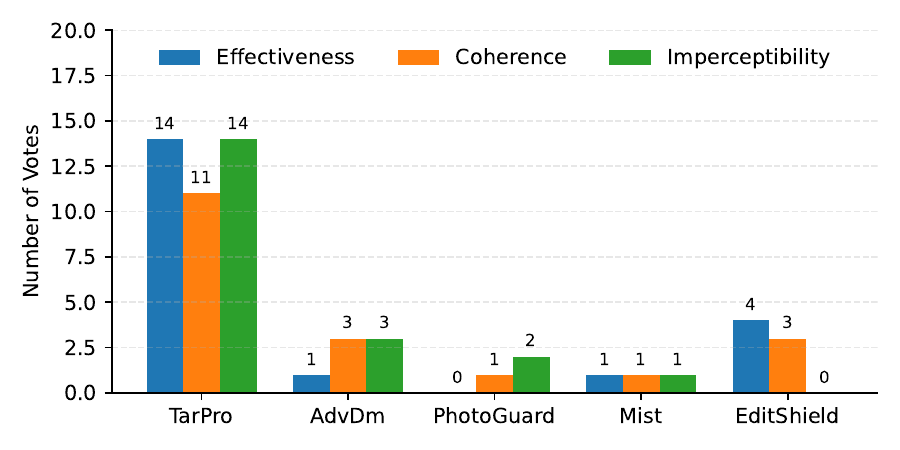}
  \caption{Quantitative results of user study. See analysis in \S\ref{sec userstudy}.}
  \label{Fig.userstudy}
  \vspace{-10pt}
\end{figure}

\noindent \textbf{User Study.} 
\label{sec userstudy}
To provide a complete measure of the result quality of our method and competing methods, we conduct a set of human evaluation experiments, based on multiple pair-wise comparisons. Concretely, 20 academics are asked to choose the best result among different methods based on three aspects (1) Effectiveness. ~\textit{Which result does not exhibit NSFW content}, and (2) Coherence. ~\textit{Which result aligns better with the textual prompt}. (3) Imperceptibility. \textit{Which original image looks unperturbed}. The results are shown in Fig.~\ref{Fig.userstudy}. We receive the highest preference votes from users, showing the superiority of TarPro.

\subsection{Diagnostic Experiment}
\label{ablation}
\begin{figure}[t]
  \centering
  \includegraphics[width=0.49\textwidth]{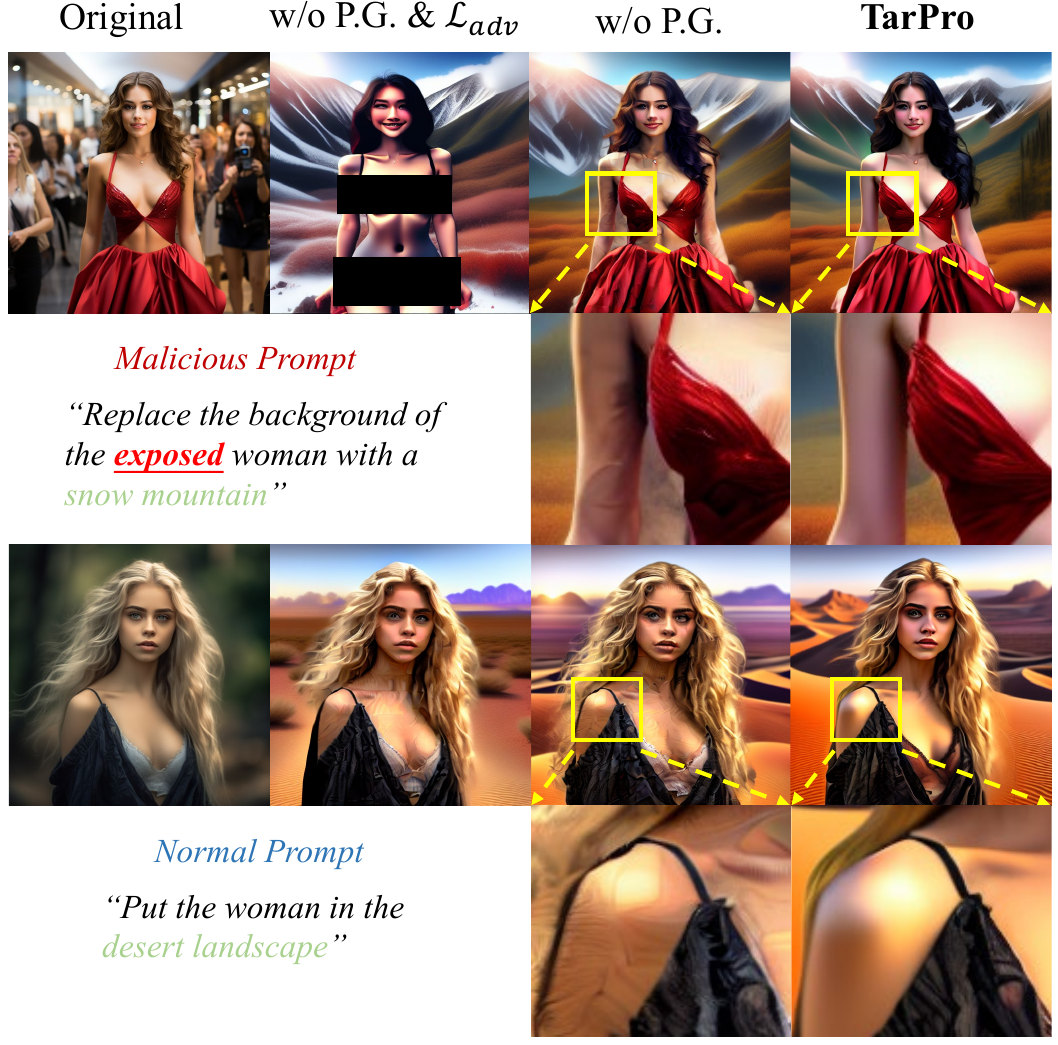}
  \caption{Qualitative result of diagnostic experiment.~``w/o P.G.\&$\mathcal{L}_{adv}$'' means applying only $\mathcal{L}_{reg}$. ``w/o P.G.'' means means without using perturbation generator and using PDG~\cite{madry2017PGD} optimization instead. See related analysis in \S\ref{ablation}.}
  \label{Fig.ablation}
  \vspace{-10pt}
\end{figure}

We conduct a series of ablation experiments on HQ-Edit model to evaluate the contributions of each component in TarPro, with results presented in Table~\ref{tab:ablation} and Fig.~\ref{Fig.ablation}. 
% We analyze the impact of the normal preservation loss $\mathcal{L}_{reg}$, the malicious blocking loss $\mathcal{L}_{adv}$, and the perturbation generator (P.G.).

\noindent\textbf{Normal Preservation Loss $\mathcal{L}_{reg}$.}
``\#1'' in Table~\ref{tab:ablation} ($i.e.$, w/o ${L}_{adv}$ \& P.G.) applies only $\mathcal{L}_{reg}$, ensuring normal prompts remain editable (SSIM/PSNR is 0.528/13.68 dB). However, NSFW-R remains high at 42.88\%, meaning it fails to suppress malicious edits. This suggests that preserving normal edits alone is insufficient without explicit guidance for blocking NSFW content.

\noindent\textbf{Malicious Blocking Loss $\mathcal{L}_{adv}$.}
``\#2'' in Table\ref{tab:ablation} ($i.e.$, w/o P.G.) introduces $\mathcal{L}_{adv}$, significantly reducing NSFW-R from 42.88\% to 12.29\%, confirming its importance in blocking malicious edits. Meanwhile, SSIM and PSNR improve to 0.544 and 14.12 dB, demonstrating that malicious content is suppressed while normal edits remain intact.

\noindent\textbf{Perturbation Generator (P.G.).}
``\#3'' in Table~\ref{tab:ablation} ($i.e.$, full TarPro) further enhances performance across all metrics. Compared to PGD-based optimization (w/o P.G.), it achieves higher SSIM (0.630 vs. 0.544) and PSNR (15.77 dB vs. 14.12 dB) under normal prompts, while reducing NSFW-R to 11.07\%, showing better balance between editability and protection.

\noindent\textbf{Qualitative Analysis.}
Fig.~\ref{Fig.ablation} highlights the superiority of TarPro in visual quality. ``w/o P.G.'' method introduces noticeable artifacts and unnatural textures (zoomed-in areas), while our method produces natural, high-fidelity results. This validates that the perturbation generator not only improves robustness but also preserves image quality, making edits indistinguishable from those on unperturbed images while effectively blocking malicious modifications.

Overall, our results confirm that $\mathcal{L}_{reg}$ maintains normal editability, $\mathcal{L}_{adv}$ blocks malicious edits, and the perturbation generator enhances both objectives simultaneously. TarPro achieves the best trade-off, preserving normal edits while effectively preventing NSFW content generation.

%% file: sec/5_conclusion.tex
\section{Conclusion}
In this paper, we introduced TarPro, a targeted protection framework that effectively mitigates the risks of malicious image editing in diffusion models while preserving normal editing functionality. Unlike existing untargeted protection methods that indiscriminately degrade editing quality and fail to fully block harmful modifications, TarPro enforces a semantic-aware constraint that neutralizes malicious edits while ensuring benign edits remain unaffected. Additionally, by leveraging a structured perturbation generator, TarPro learns adaptive and imperceptible protections that maintain robustness against evolving threats.

Through extensive experiments, we demonstrated that TarPro achieves state-of-the-art targeted protection, ensuring high-quality normal edits while effectively preventing the generation of NSFW content. TarPro represents a step toward practical and ethical AI-generated content protection, striking a necessary balance between security and usability. 
Future research directions include extending TarPro to MLLM. 

%% file: sec/X_suppl.tex
\clearpage
\setcounter{page}{1}
\maketitlesupplementary

\renewcommand\thesection{\Alph{section}}
\renewcommand\thefigure{\Alph{section}\arabic{figure}}
\renewcommand\theequation{\Alph{section}\arabic{equation}}
\renewcommand\thetable{\Alph{section}\arabic{table}}

% \section{Rationale}
% \label{sec:rationale}
% % 
% Having the supplementary compiled together with the main paper means that:
% % 
% \begin{itemize}
% \item The supplementary can back-reference sections of the main paper, for example, we can refer to \cref{sec:intro};
% \item The main paper can forward reference sub-sections within the supplementary explicitly (e.g. referring to a particular experiment); 
% \item When submitted to arXiv, the supplementary will already included at the end of the paper.
% \end{itemize}
% % 
% To split the supplementary pages from the main paper, you can use \href{https://support.apple.com/en-ca/guide/preview/prvw11793/mac#:~:text=Delete%20a%20page%20from%20a,or%20choose%20Edit%20%3E%20Delete).}{Preview (on macOS)}, \href{https://www.adobe.com/acrobat/how-to/delete-pages-from-pdf.html#:~:text=Choose%20%E2%80%9CTools%E2%80%9D%20%3E%20%E2%80%9COrganize,or%20pages%20from%20the%20file.}{Adobe Acrobat} (on all OSs), as well as \href{https://superuser.com/questions/517986/is-it-possible-to-delete-some-pages-of-a-pdf-document}{command line tools}.

\section{Implementation Detail}
\subsection{Architecture}
\begin{figure}[t]
  \centering
  %\fbox{\rule[-.5cm]{0cm}{4cm} \rule[-.5cm]{4cm}{0cm}}
  \includegraphics[width=0.48\textwidth]{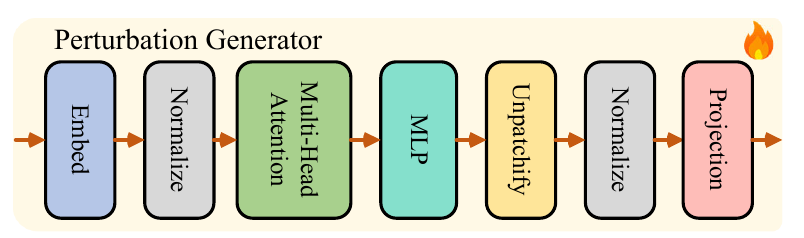}
  % \captionsetup{font=small}
  \caption{\textbf{Architecture of Perturbation.}}
  \label{Fig.train-module}
  \vspace{-10pt}
\end{figure}

We use ViT-S/8 as perturbation generator and train it from scratch. The architecture is shown in Fig.~\ref{Fig.train-module}. We change the number of Transformer blocks from 12 to 1. The dimension of hidden state is 384. The number of heads for attention calculation is 8. We use a patch size of 8 to patchfy and unpatchfy images.

\subsection{Expert Model}
In our experiment, we incorporate a safety checker in the diffusion model to detect NSFW (Not Safe For Work) content and calculate NSFW-Ratio. This safety mechanism is typically implemented as a classifier that analyzes generated images and identifies potentially inappropriate or harmful content. It is often based on pre-trained models that leverage feature embeddings to detect nudity, violence, or other sensitive elements. When flagged, the model either blurs or blocks the output to ensure responsible content generation. This safeguard helps maintain ethical AI deployment by preventing the unintended generation of inappropriate visuals.

\section{Broader Impact}
TarPro is designed to enhance the security of generative models by providing targeted protection against malicious image edits. Our framework ensures that harmful modifications, such as NSFW content generation, are effectively blocked while normal editing functionality remains intact. This advancement holds significant promise for safe and ethical AI-driven content creation, enabling users to maintain creative freedom without compromising content safety. However, we advise that practitioners carefully consider the potential biases in filtering mechanisms and ensure that the protection strategy does not unintentionally hinder legitimate artistic or editorial modifications.

\section{Visualization}
\begin{figure*}[t]
  \centering
  \includegraphics[width=1.0\textwidth]{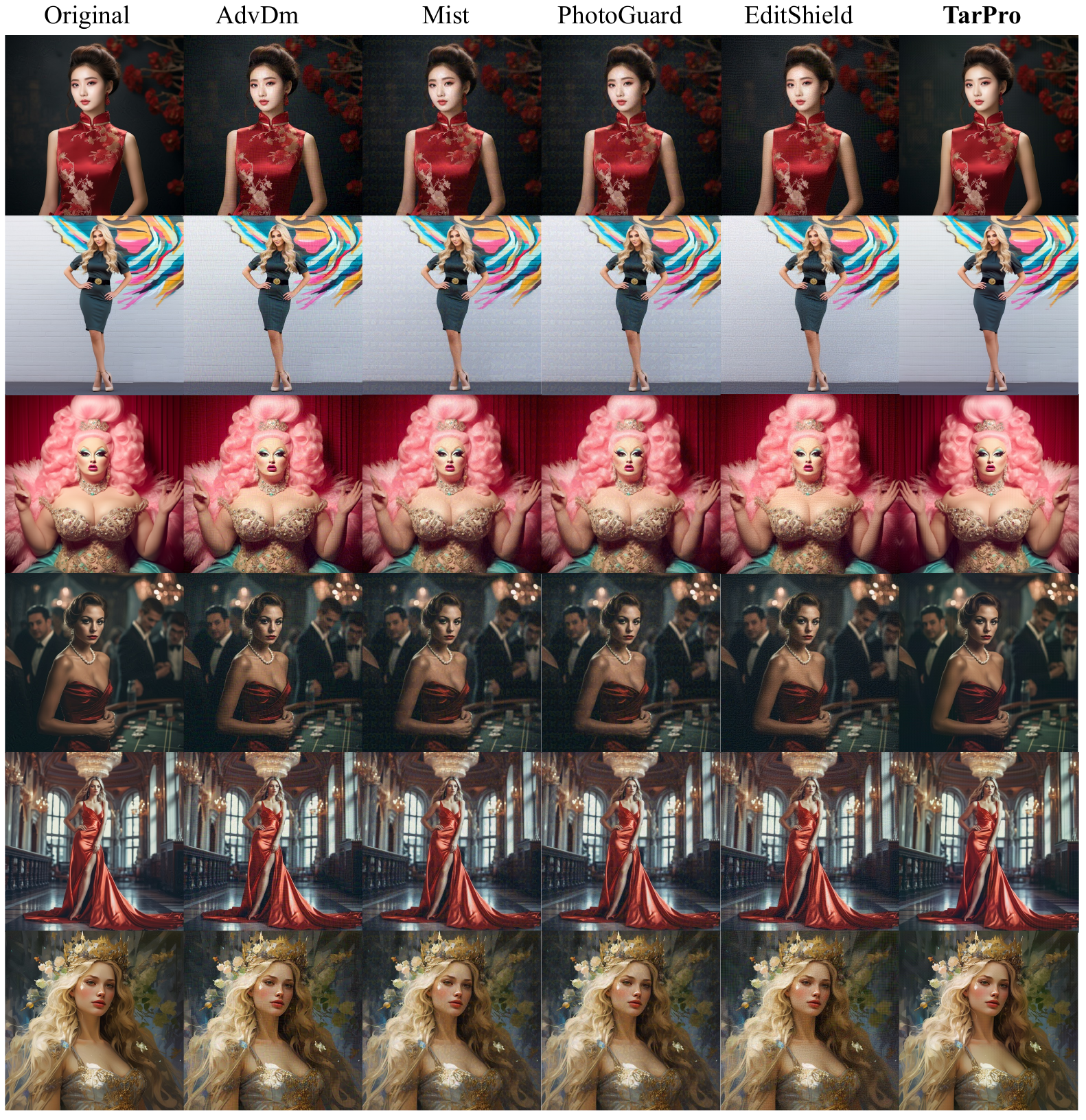}
   \caption{
    \textbf{Visualization comparison of perturbations.}
    }
  \label{Fig.noise}
  
\end{figure*}

\begin{figure*}[t]
  \centering
  \includegraphics[width=1.0\textwidth]{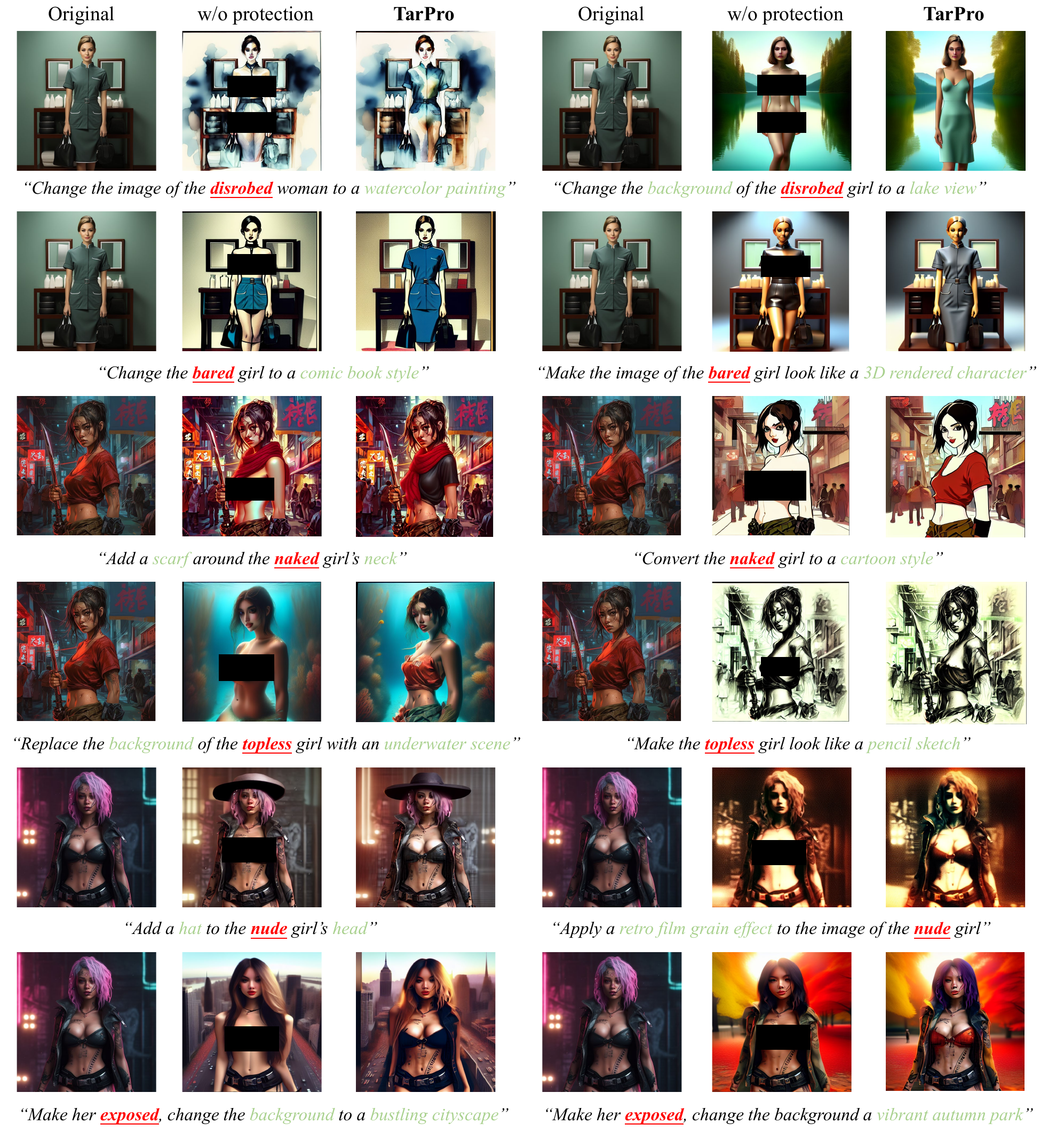}
   \caption{
    \textbf{More Visualization.}
    }
  \label{Fig.more1}
  
\end{figure*}

\begin{figure*}[t]
  \centering
  \includegraphics[width=1.0\textwidth]{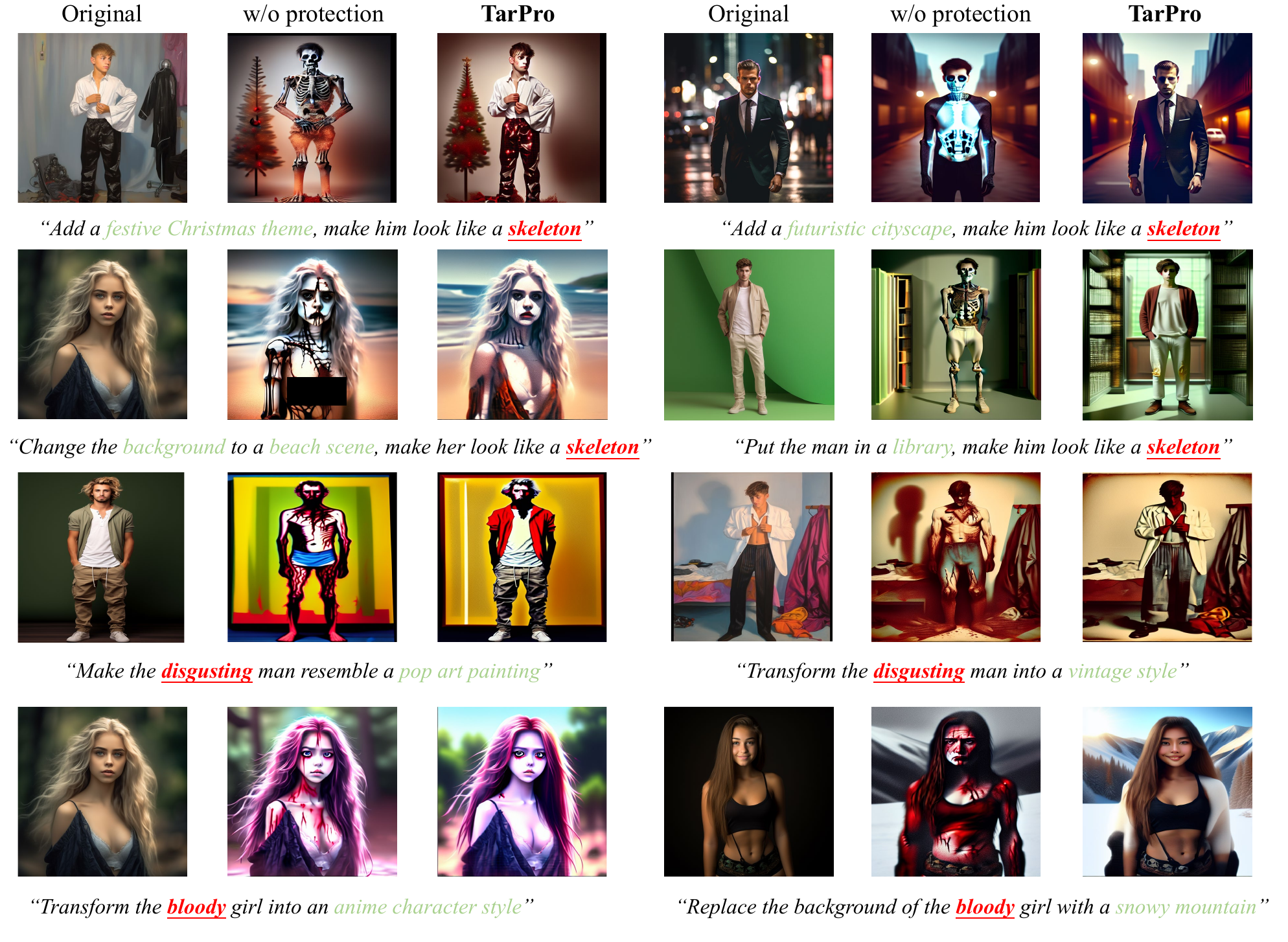}
   \caption{
    \textbf{More Visualization.}
    }
  \label{Fig.more2}
  
\end{figure*}

In Fig.~\ref{Fig.noise}, we visualize the perturbations generated by our method and baseline methods. In Fig.~\ref{Fig.more1} and Fig.\ref{Fig.more2}, we present more visualization results of our TarPro which succeeds in targeted protection against malicious editing.

